\documentclass{article}

\usepackage[final]{nips_2016}

\usepackage[utf8]{inputenc} 
\usepackage[T1]{fontenc}    
\usepackage{hyperref}       
\usepackage{url}            
\usepackage{booktabs}       
\usepackage{amsfonts}       
\usepackage{nicefrac}       
\usepackage{microtype}      
\usepackage{graphicx}
\usepackage{common}
\usepackage{subfigure}
\title{Dilated Convolutions for Modeling \\ Long-Distance Genomic Dependencies}

\author{
  Ankit Gupta \\
  School of Engineering and Applied Sciences \\
  Harvard University\\
  Cambridge, MA 02138 \\
  \texttt{ankitgupta@college.harvard.edu} \\
  \And
  Alexander M. Rush \\
  School of Engineering and Applied Sciences \\
  Harvard University\\
  Cambridge, MA 02138 \\
  \texttt{srush@seas.harvard.edu} \\
}

\begin{document}

\maketitle

\begin{abstract}
  We consider the task of detecting regulatory elements in the human
  genome directly from raw DNA. Past work has focused on small
  snippets of DNA, making it difficult to model long-distance
  dependencies that arise from DNA's 3-dimensional conformation. In
  order to study long-distance dependencies, we develop and release a
  novel dataset for a larger-context modeling task. Using this new
  data set we model long-distance interactions using dilated
  convolutional neural networks, and compare them to standard
  convolutions and recurrent neural networks. We show that dilated
  convolutions are effective at modeling the locations of regulatory
  markers in the human genome, such as transcription factor binding
  sites, histone modifications, and DNAse hypersensitivity
  sites. 
\end{abstract}

\section{Introduction}
\label{intro}


Gene expression is controlled by a variety of \textit{regulatory
  factors} that determine which genes are expressed in which
environmental conditions \citep{perkins2005expanding}.  Due to
proteins called histones that DNA winds around, parts of DNA are more
accessible to binding than others, and so DNA accessibility is a
regulatory factor. Modifications to histones can affect the
conformation of DNA, so histone modifications are also regulatory
factors. Furthermore, proteins that bind to DNA and affect
transcriptional activity are called transcription factors and are
also regulatory factors. The activity that they affect can be
thousands of base pairs away \citep{blackwood1998going}.

These interactions imply that nucleotides far apart in a
1-dimensional DNA sequence may interact in its 3-dimensional
conformation, and so expression is governed by both local and
long-distance dependencies. As a result, it may be important to
incorporate DNA regions that are far away in 1-D space when modeling
regulatory markers.  However, past data sets for binding site
prediction have used small snippets of DNA
\citep{deepbind,deepsea,danq}, which limits the ability to model these
interactions.

This work addresses this question by introducing a new dataset for
this problem allowing for long-distance interactions, and a new model
using dilated convolutions to predict the locations of regulatory
markers.  We learn a mapping from a DNA region, specified as a
sequence of nucleotides, to the locations of regulatory markers in
that region. Dilated convolutions can capture a hierarchical
representation of a much larger input space than standard
convolutions, allowing them to scale to large context sizes. We
compare dilated convolutions to other modeling methods from deep
learning and past work, including standard convolutions and recurrent neural
networks, and show that they present an advancement over existing
models.  All code, data, and scripts for these experiments are available at
\url{https://github.com/harvardnlp/regulatory-prediction}.

\begin{figure*}[h!]
\centering     
\subfigure[Convolution]{\label{fig:conv-receptive-field-a}\includegraphics[width=
.48\columnwidth]{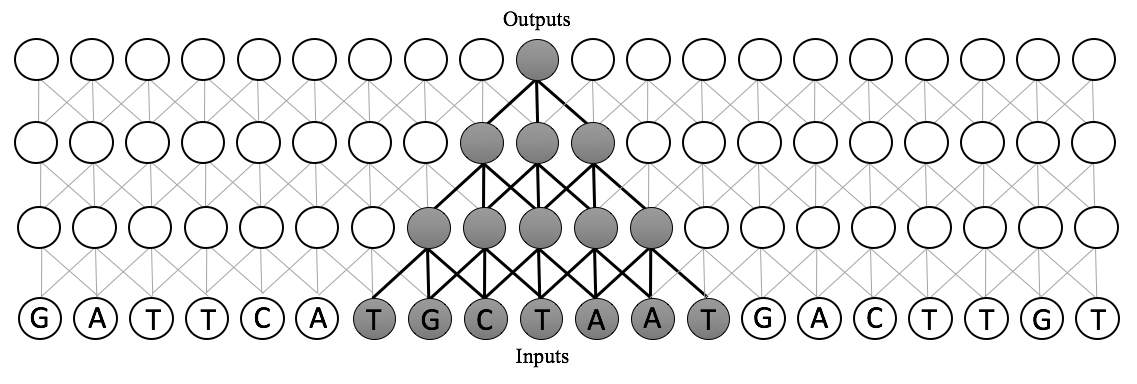}}
\subfigure[Bidirectional LSTM]{\label{fig:lstm_receptive_field}\includegraphics[width=.48\columnwidth]{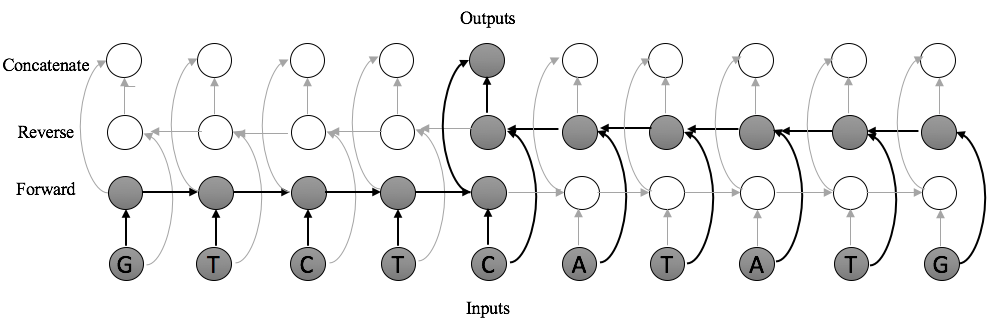}}
\subfigure[Dilated Convolution]{\label{fig:conv-receptive-field-b}\includegraphics[width=.5\columnwidth]{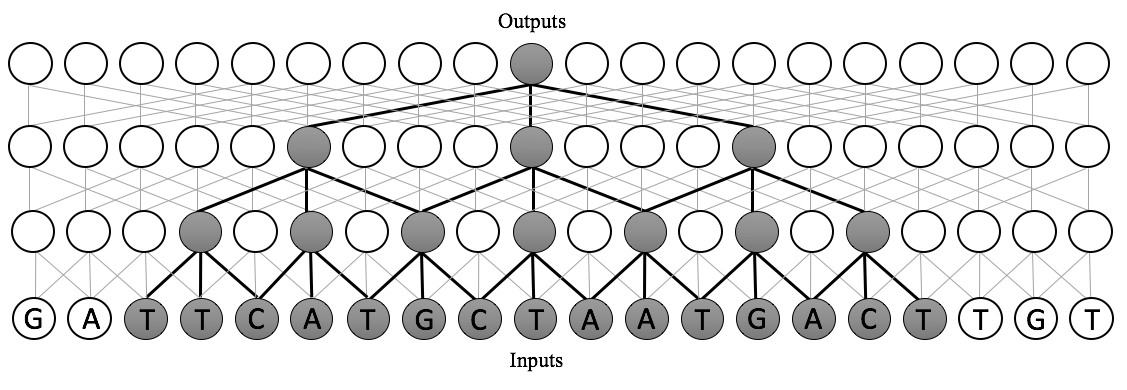}}\\
\caption[Convolution Receptive Fields]{\small{Visual representation of the models. All take a sequence of nucleotides as input. The receptive field is shaded. The convolution has a short path between inputs and predictions but a small receptive field. The Bi-LSTM has a large receptive field, the whole sequence, but may have a long path between a nucleotide and prediction. The dilated convolution has both a short path from the input \textit{and} a large receptive field. }}
\label{fig:comparison_diagram}
\end{figure*}
\section{Models}
\vspace*{-2mm}
\paragraph{Convolutional and Recurrent Neural Networks}
The two most widely applied models for sequence tasks are
convolutional neural networks (CNNs) and recurrent neural networks
(RNNs). CNNs consist of a series of layers each applying a linear
transformation followed by a non-linearity to a sliding kernel window of
inputs \citep{lecun1995convolutional}, shown in Figure
\ref{fig:conv-receptive-field-a}. RNNs apply the same to each
element of a sequence in time. An RNN with state size $M$
processes a sequence of inputs $\{\boldx_i\}_N$ by updating a
state vector $\boldc_i \in \mathbb{R}^M$ for each timestep $i$ such that
$\boldc_{i+1} = R(\boldc_{i}, \boldx_i)$.
Concatenating a forward and reverse RNN gives a bidirectional
RNN, as in Figure \ref{fig:lstm_receptive_field}.

The interactions in a network can be quantified by the
receptive field. Formally the \textit{receptive field} of a node is
the subset $\mathcal{R}$ of input elements $\{\boldx_i\}$ that can
impact its value \citep{yu2015multi}. For CNNs, the size of the
receptive field is linear in the number of layers and the kernel
width. Thus, scaling the receptive field to incorporate a large input
introduces more layers, making training more difficult. Bidirectional RNNs on the
other hand have a receptive field of the whole input, but require
gradients to travel long-distances over time. Long Short-Term Memory
(LSTM) cells \citep{lstm} circumvent some of these issues by using
trainable gated connections that allow the cell state to propagate
more efficiently. However, with very long sequences, as is the case in
genetic data, LSTMs still have trouble learning very long-distance relationships.

\paragraph{Dilated Convolutions}
\label{section:dilated_conv}
Dilated convolutions \citep{yu2015multi} offer a middle
ground between these two models with wide receptive fields and short
distance gradient propagation. In CNNs each
kernel window consists of adjacent inputs, while dilated convolutions
introduce gaps (dilation) between inputs.  With dilation $d$, the
window starting at location $i$ of size $k$ is
\begin{align*} \begin{bmatrix} \boldx_i & \boldx_{i + d} & \boldx_{i + 2d} & \cdots & \boldx_{i + (k-1)\cdot d}
\end{bmatrix} \end{align*}
\citet{yu2015multi} show that by stacking these convolutions with
increasingly large $d$, we can expand the receptive field of each
output exponentially. This allows them to have large receptive fields, but still short backpropagations, as shown in Figure \ref{fig:conv-receptive-field-b}. We take
advantage of this structure when modeling genetic regulation. Dilated convolutions have been used for image segmentation \citep{yu2015multi}, text-to-speech \citep{wavenet}, and text classification \citep{strubell2017fast}.
\section{Experiments}
\vspace*{-2mm}
\paragraph{Dataset 1: Short Sequence Prediction Benchmark \citep{deepsea}}
As a preliminary experiment, we test dilated convolutions on a standard benchmark task and compare them to CNNs and LSTMs. We use the \citet{deepsea} dataset to predict the presence of regulatory markers in short DNA sequences. Each input is a vector $\boldx \in \mcV^{d}$, where
$d = 1000$ is the sequence length, and $\mcV = \{A, C, T, G\}$. Each output vector
$\boldy \in \{0, 1\}^{k}$ indicates whether each of the $k = 919$ regulatory markers is present in the middle
200bp of $\boldx$. The markers include TFBSs,
histone modifications, and DNAse hypersensitivity sites (accessible DNA regions).

We train several architectures and report
the mean PR AUC scores for each category in Table
\ref{tab:reimp_results}. The models in this task have a final
fully-connected layer, which implies that the receptive field of every
output contains the whole input (1000 bp). The differences between models is
explained by the extent to which each captures meaning 
prior to the final layer. Note that there was dropout
\citep{srivastava2014dropout} and batch normalization
\citep{ioffe2015batch} between every layer, and we select
hyperparameters using grid search. Details are given in
the Supplemental Information section.

\begin{table}[t]
\centering
\begin{tabular}{lccccccccc}
\toprule
&  \multicolumn{7}{c}{Dataset 1: Short Sequence Prediction Task } \\
\midrule
 Model &  & Hidden & Type & Params (Best-Case) & \multicolumn{3}{c}{Test Set PR AUC } \\
  &  & & & & TFBS & Hist & DNAse \\
 \midrule
 \textsc{LR} & & 0  & - & 3676919&0.042 & 0.143 &  0.097 \\
 \textsc{FF} & & 1  & Fully &4551919&0.046 & 0.181 & 0.106 \\
 \textsc{CNN3} & &3 &CNN &155159839& 0.205 & 0.273 &0.319 \\
 \textsc{Dilated3} & &  3 &Dilated &37056519&0.255  & 0.293  &  0.372  \\
 \textsc{Dilated6} & & 6  &Dilated &25758079& 0.285  & 0.320 & 0.396\\
 \textsc{Bi-LSTM} & & 2 &Bi-LSTM &46926479&  \textbf{0.305} & \textbf{0.340} & \textbf{0.407}  \\
 \bottomrule
\end{tabular}
\caption{\small{\label{tab:reimp_results}  Models and Test Set Precision-Recall Area Under Curve (PR AUC) scores for Dataset 1. We report the number of parameters in the hyperparameter configuration selected using grid search with a held-out validation set. \textsc{CNN3} is the model from \citet{deepsea}, and Bi-LSTM is the bidirectional LSTM model from \citet{danq}. Our \textsc{Dilated6} model performs better than the standard convolutions on all three types of predictions and only slightly underperforms the bidirectional LSTM model. All scores are based on reimplementations. Model size varied across hyperparameter configurations. }}
\end{table}
Notably the \textsc{Dilated6} model performs much better than the \textsc{CNN3} model from \citet{deepsea}, and approaches the performance of the \textsc{Bi-LSTM} model.
The \textsc{Bi-LSTM} is the most effective
model, which is reasonable since the sequences are short. This gives a proof-of-concept that dilated convolutions can capture nucleotide structure better than standard
convolutions.
\vspace*{-2mm}
\paragraph{Dataset 2: Complete Genome Labeling with Long-Range Inputs  }
To model long-distance dependencies, we introduce a new dataset that
models the problem as long-distance sequence labeling instead of
prediction. In particular: (1) the input sequences are longer, with
each a $d = 25000$ bp sequence, and (2) the outputs annotate at
nucleotide-resolution, making this a dense labeling task. We predict
the presence of all regulatory markers at each nucleotide, rather than
per 200bp.

We use hg19 to extract input sequences
and $k=919$ regulatory marker locations from ENCODE \citep{encode2012integrated}. Thus, we get pairs of input and output sequences $\{(\boldx,
\boldy)\}$ with $\boldx \in \mcV^{d}$ and $\boldy \in \{0, 1\}^{d \times k}$. The inputs have vocabulary
$\mcV = \{N, A, C, T, G\}$, where
the N character represents nucleotides with high uncertainty. We remove sequences with >10\% unknown
nucleotides or with >10\% part of a multi-mapped sequence, meaning part of a region
that maps to several genomic locations. This left $n = 93880$ non-overlapping sequences that were $d =
25000$ in length, totaling 2.3 billion nucleotides. We train on 80\% of the data, and split the rest into validation and test sets.

We train models representing the various architectures on this new dataset. We showed above with Dataset 1 that LSTM-based models were the most effective at predicting regulatory marker locations when given only a small number of nucleotides, and we can now study their relative success when given more context. Since this is a base-pair level prediction task, we no longer have a fully-connected layer for each model. We summarize the models for this task and report PR AUC scores in Table \ref{tab:results_validation_imp}. We report the best models across all hyperparameters, including number of filters, dropout, learning rate, batch norm decay, and hidden layer size, using grid search on a held-out validation set. For the dilated convolution, we use dilations of 1, 3, 9, 27, and 81 in sequence.

\begin{table*}[t]
\centering
\begin{tabular}{lccccccccc}
 \toprule
& &\multicolumn{7}{c}{Dataset 2: Complete Genome Labeling } \\
\midrule

 Model & Layers & Type& Params & \multicolumn{3}{c}{Validation PR AUC } & \multicolumn{3}{c}{Test PR AUC } \\
  &   &&& TFBS & Hist & DNAse & TFBS & Hist & DNAse \\
 \midrule
 \textsc{CNN1} & 1 &CNN&137187 & 0.013 & 0.053 & 0.035 & - & - & - \\
 \textsc{CNN7}  & 3&CNN& 341803& 0.059 & 0.115 &  0.100 &- &- & - \\
 \textsc{CNN7} & 7&CNN& 656363 & 0.167 & 0.166  & 0.180 & 0.167  &0.165  & \textbf{0.186} \\
 \textsc{ID-CNN} &15 &Dilated& 635739 & 0.166 &  0.247 & 0.147 & 0.171 & 0.236 & 0.142 \\
 \textsc{Bi-LSTM} & 4&Bi-LSTM& 764395 &  0.104 & 0.288  & 0.116 & 0.107  & 0.264 & 0.113 \\
 \textsc{Dilated} &6 &Dilated& 631263 &0.274  &  0.279 & 0.178 & \textbf{0.274}  & \textbf{0.273} &0.179 \\
 \bottomrule
\end{tabular}
\caption{\small{\label{tab:results_validation_imp} Models for Dataset 2 and Precision-Recall Area Under Curve (PR AUC) scores. We report the number of parameters in the best-performing hyperparameter configuration. ID-CNN is the model from \citet{strubell2017fast}. We see much higher
performance using dilated convolutions on predicting transcription factor binding sites and histone modifications, but no
improvement on predicting DNAse hypersensitivity sites. Note that this because this is a new dense prediction task, these results should not be directly compared to those in Table \ref{tab:reimp_results}.}}
\end{table*}

Dilated convolutional models perform the best on both TFBS
and histone modification prediction, and do marginally worse than the best non-dilated
models on predicting DNAse hypersensitivity sites. This shows that dilated convolutions can be effective at capturing
structure in the genome. In contrast, while \textsc{Bi-LSTM} performed better than \textsc{CNN3}, it did worse
than the convolutions, particularly on TFBS prediction. This suggests that the LSTM architecture is less effective at this task, either due to vanishing gradients or difficulty in learning gated recurrent connections. This suggests that even though LSTMs are effective on short sequences, when trying to capture properties of long genetic sequences, dilated convolutions are an important architecture to consider.

On DNAse hypersensitivity prediction, standard convolutions do well. This may be because accessible regions locally have highly explanatory motifs and having additional context far away does not improve the accuracy, while the other two marker types are more easily characterized with access to distal motifs. This is also consistent with the high performance of both \textsc{Dilated6} and Bi-LSTM in predicting DNAse sites in Dataset 1 compared to TFBS and histone modifications.

\begin{figure*}[h]
\centering     
\subfigure[\textsc{CNN7}]{\label{fig:result_receptive-field-CNN7}\includegraphics[width=.2\columnwidth]{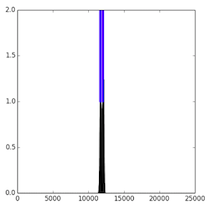}}
\subfigure[\textsc{Dilated}]{\label{fig:result-receptive-field-dilated}\includegraphics[width=.21\columnwidth]{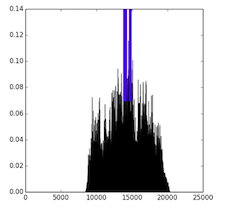}}
\subfigure[\textsc{Bi-LSTM}]{\label{fig:result-receptive-field-lstm}\includegraphics[width=.2\columnwidth]{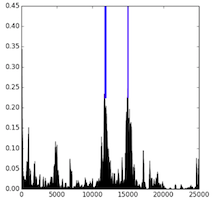}}
\caption[Task 2 Receptive Fields]{\small{We visualize the norm of the gradient to the
inputs (black). This gives an indication about the actual receptive field that was used to make a decision at
the outputs (blue). Notably, we see that \textsc{CNN7} has a narrow receptive field, \textsc{Dilated} has a wide high-magnitude field, and \textsc{Bi-LSTM} has a wide but low-magnitude field.}}
\label{fig:result_receptive_field}
\end{figure*}

To further investigate the trained models, we visualize receptive fields by sampling a validation sequence and backpropagating
an error of 1 from every positive output for a random
regulatory marker. In Figure \ref{fig:result_receptive_field}, we plot the
output locations (blue) and the norm
of the error to the inputs (black), which gives a visual representation of the receptive field. We observe that the standard convolution has a narrow receptive field while the
dilated convolution has a wider one, as the gradient is high for a wider input in Figure \ref{fig:result-receptive-field-dilated}. In contrast, the LSTM model in Figure
\ref{fig:result-receptive-field-lstm} has gradient backpropagated widely, but it usually has a low magnitude. It does not appear that the LSTM models are able to learn the
long-distance dependencies that the dilated model captures, meaning that though LSTM models were successful on short inputs in Dataset 1, they were unable to scale to larger inputs on Dataset 2.
\vspace*{-4mm}
\section{Conclusion}
\vspace*{-2mm}
We introduce a new data set with larger DNA contexts and base pair
level data. On this data set, we show that dilated convolutions can
outperform both CNNs and LSTMs and appear to capture long-distance
relationships in DNA. This suggests that dilated convolutions are an important architecture to consider for genetic modeling. Next, we intend to incorporate more DNA
structural information, such as Hi-C data that measures DNA
conformation \citep{belton2012hi}. We also intend to study whether a
hybrid architecture can effectively predict all marker types, particularly the DNAse sites.

\section{References}

 \bibliography{icml_compbio}

\begin{thebibliography}{}

\bibitem[Alipanahi et~al., 2015]{deepbind}
Alipanahi, B., Delong, A., Weirauch, M.~T., and Frey, B.~J. (2015).
\newblock Predicting the sequence specificities of dna-and rna-binding proteins
  by deep learning.
\newblock {\em Nature biotechnology}.

\bibitem[Belton et~al., 2012]{belton2012hi}
Belton, J.-M., McCord, R.~P., Gibcus, J.~H., Naumova, N., Zhan, Y., and Dekker,
  J. (2012).
\newblock Hi--c: a comprehensive technique to capture the conformation of
  genomes.
\newblock {\em Methods}, 58(3):268--276.

\bibitem[Blackwood and Kadonaga, 1998]{blackwood1998going}
Blackwood, E.~M. and Kadonaga, J.~T. (1998).
\newblock Going the distance: a current view of enhancer action.
\newblock {\em Science}, 281(5373):60--63.

\bibitem[Consortium et~al., 2012]{encode2012integrated}
Consortium, E.~P. et~al. (2012).
\newblock An integrated encyclopedia of dna elements in the human genome.
\newblock {\em Nature}, 489(7414):57--74.

\bibitem[Hochreiter and Schmidhuber, 1997]{lstm}
Hochreiter, S. and Schmidhuber, J. (1997).
\newblock Long short-term memory.
\newblock {\em Neural computation}, 9(8):1735--1780.

\bibitem[Ioffe and Szegedy, 2015]{ioffe2015batch}
Ioffe, S. and Szegedy, C. (2015).
\newblock Batch normalization: Accelerating deep network training by reducing
  internal covariate shift.
\newblock {\em arXiv preprint arXiv:1502.03167}.

\bibitem[LeCun et~al., 1995]{lecun1995convolutional}
LeCun, Y., Bengio, Y., et~al. (1995).
\newblock Convolutional networks for images, speech, and time series.
\newblock {\em The handbook of brain theory and neural networks},
  3361(10):1995.

\bibitem[Oord et~al., 2016]{wavenet}
Oord, A. v.~d., Dieleman, S., Zen, H., Simonyan, K., Vinyals, O., Graves, A.,
  Kalchbrenner, N., Senior, A., and Kavukcuoglu, K. (2016).
\newblock Wavenet: A generative model for raw audio.
\newblock {\em arXiv preprint arXiv:1609.03499}.

\bibitem[Perkins et~al., 2005]{perkins2005expanding}
Perkins, D.~O., Jeffries, C., and Sullivan, P. (2005).
\newblock Expanding the ``central dogma'': the regulatory role of nonprotein
  coding genes and implications for the genetic liability to schizophrenia.

\bibitem[Quang and Xie, 2016]{danq}
Quang, D. and Xie, X. (2016).
\newblock Danq: a hybrid convolutional and recurrent deep neural network for
  quantifying the function of dna sequences.
\newblock {\em Nucleic acids research}, page gkw226.

\bibitem[Srivastava et~al., 2014]{srivastava2014dropout}
Srivastava, N., Hinton, G.~E., Krizhevsky, A., Sutskever, I., and
  Salakhutdinov, R. (2014).
\newblock Dropout: a simple way to prevent neural networks from overfitting.
\newblock {\em Journal of Machine Learning Research}, 15(1):1929--1958.

\bibitem[Strubell et~al., 2017]{strubell2017fast}
Strubell, E., Verga, P., Belanger, D., and McCallum, A. (2017).
\newblock Fast and accurate sequence labeling with iterated dilated
  convolutions.
\newblock {\em arXiv preprint arXiv:1702.02098}.

\bibitem[Yu and Koltun, 2015]{yu2015multi}
Yu, F. and Koltun, V. (2015).
\newblock Multi-scale context aggregation by dilated convolutions.
\newblock {\em arXiv preprint arXiv:1511.07122}.

\bibitem[Zhou and Troyanskaya, 2015]{deepsea}
Zhou, J. and Troyanskaya, O.~G. (2015).
\newblock Predicting effects of noncoding variants with deep learning-based
  sequence model.
\newblock {\em Nature methods}, 12(10):931--934.

\end{thebibliography}
\bibliographystyle{apalike}

\section{Acknowledgements}
We thank David Kelley for his insights on the biological mechanisms behind these regulatory factors, the necessary steps to pre-process ENCODE data, and his general guidance in interpreting the results of this work.

\newpage
\section{Supplemental Information: Architecture Details and Hyperparameters}
\begin{figure}[h]
\centering     
\includegraphics[width=\columnwidth]{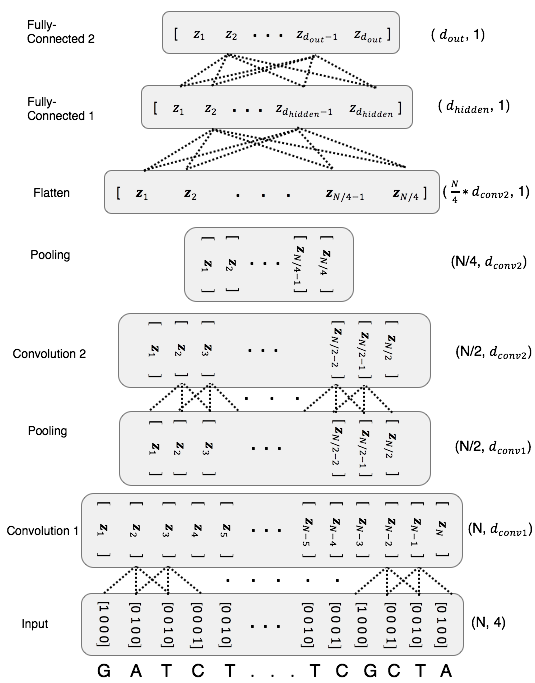}
\caption[Dataset 1 Architecture]{An overview of the architecture we use on Dataset 1. The input (bottom) is a 
sequence of N nucleotides of DNA, represented as one-hot encoded vectors. The 
shape of the tensor at each layer is listed on the right. This example shows 2 
convolutions, each followed by a pooling layer, and finally two fully-connected layers.
Note that there are more convolutions and pooling in some of our models, and we also use dilated 
convolutions. Additionally, we use a bidirectional LSTM
in between the Input and Flatten layers for one of the baseline models. The fully 
connected layers apply to the flattened input. There are $d_{out} = 919$ outputs 
for each sequence of $N = 1000$ inputs.}
\label{fig:exp1_model}
\end{figure}

\paragraph{Dataset 1: Short Sequence Prediction Benchmark}
We summarize the high-level architecture that we use for this task in Figure \ref{fig:exp1_model} with an example with two 
convolutional layers. All of the models have the same input representation, and have fully-connected layers at the end. The 
differences between the models exists between the Input 
and Flatten layers of Figure \ref{fig:exp1_model}, as each model uses a different type of convolutional layer 
and one uses a bidirectional LSTM. Note that in the logistic regression and multilayer perceptron models, we directly 
flatten the inputs and use fully-connected layers, rather than first applying convolutions. 

We list the hyperparameters we tested in Table \ref{tab:exp1_hyperparameters} in finding the optimum \textsc{Dilated6} model. We adjusted the number of filters in each layer, represented by $\{f_1, f_2, f_3\}$. $f_1$ and $f_2$ are for the first two layers and $f_3$ is for all further layers of the model. We also test the kernel width of the convolution, represented by $k$. Finally, we adjust the size of the final hidden layer, represented by $d_{hidden}$.

\begin{table}[h]
\centering
\begin{tabular}{cc}
 \toprule
 Hyperparameter & Options  \\
 \midrule
Learning Rate & Between .0001 and .1 \\
Batch Norm Decay Rate & \{.9, .99\} \\
Dropout & \{.1, .01\} \\
k & \{4, 8, 12\} \\
$f_1$ & \{120, 240, 360\} \\
$f_2$ &\{120, 240, 360\} \\
$f_3$ &\{120, 240, 360\} \\
$d_{hidden}$ & \{500, 919, 1500\}\\ 
 \bottomrule
\end{tabular}
\caption{\label{tab:exp1_hyperparameters} Hyperparameters tested in Task 1.}
\end{table}
\noindent

\begin{figure}
\centering     
\includegraphics[width=\columnwidth]{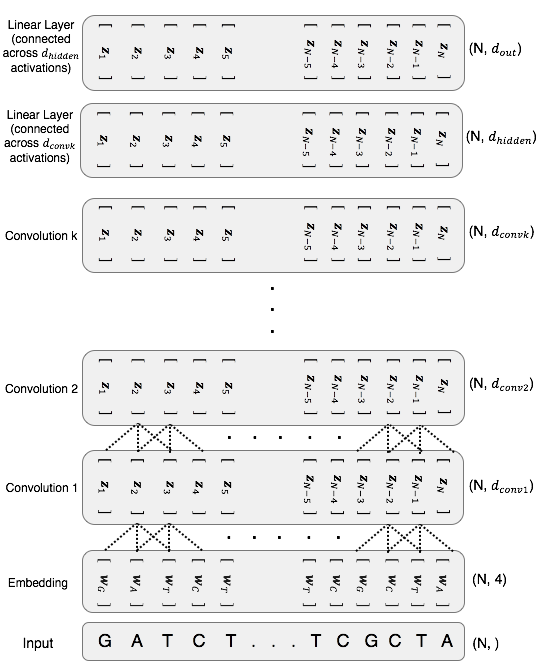}
\caption[Dataset 2 Architecture]{An overview of the convolutional architecture used on Dataset 2. 
The input (bottom) is a sequence of N nucleotides. The shape of the tensor at 
each layer is listed on the right. The first layer embeds each element 
into a higher-dimensional vector space. After that, there are $k$ convolutions (or dilated convolutions) with
ReLU activations. Furthermore, we test adding pooling layers with a stride of 1
in between convolutional layers. In the LSTM model, the convolutions
are replaced with a bidirectional LSTM. Finally, the last two linear 
layers do not connect elements that are horizontally separated in the above 
diagram, but only those within the activations for a particular element. So, 
they apply a linear map to each of the activations $\mathbf{z}_i \in 
\mathbb{R}^{d_{convk}}$. In other words, the first linear layer has 
weights which are $\boldW^{(1)} \in\mathbb{R}^{d_{convk} \times d_{hidden}}$, and 
the second has weights which are $\boldW^{(2)} \in\mathbb{R}^{d_{hidden} \times 
d_{out}}$. In practice, $N = 25000$, and $d_{out} = 919$.}
\label{fig:exp2_model}
\end{figure}

\paragraph{Dataset 2: Complete Genome Labeling with Long-Range Inputs  }
We summarize the high-level architecture of this task in Figure \ref{fig:exp2_model}. The differences 
between the models we implement for this task exists between the 
Embedding and first Linear (fully-connected) layer of Figure 
\ref{fig:exp2_model}. In contrast with Dataset 1, we are now making dense predictions,
meaning one set of predictions for every nucleotide in the input sequence, which leads to a different
model structure. 

We list the hyperparameters we tested in Table \ref{tab:exp2_hyperparameters}. For the convolutions, we adjusted the number of filters in each layer, represented by $\{f_1, f_2, f_3\}$. $f_1$ and $f_2$ are for the first two layers and $f_3$ is for all further layers of the model, if applicable. For the LSTM, $f_1$ is the number of filters in the first convolution, $f_2$ referred to the state size of each LSTM, and $f_3$ is the number of filters in the deconvolution. For the LSTM, we also adjust the stride for the first convolution and final deconvolution. Each model had one hidden layer after the convolutions, with $d_{hidden}$ units.
\begin{table}[h!]
\centering
\begin{tabular}{ccc}
 \toprule
 Hyperparameter & Conv Options & LSTM Options  \\
 \midrule
Learning Rate & [.001, .1]   &  [.001 and .1]   \\
Batch Decay & \{.9, .99\}& \{.9, .99\} \\
Dropout &\{.1, .01\} & \{.1, .01\}\\
$f_1$ & \{ 64, 128 \} & 128\\
$f_2$ & \{120, 240\} & \{20, 40, 80\} \\
$f_3$ & \{50, 128\}  & \{50, 128\} \\
Stride & 10 & \{10, 20, 50, 100\} \\
$d_{hidden}$ & \{125, 200\} & \{64, 125, 200\}\\
 \bottomrule
\end{tabular}
\caption{\label{tab:exp2_hyperparameters} Hyperparameters tested in Task 2.}
\end{table}

\end{document}